\documentclass[12pt,a4paper]{article}
\usepackage{amsmath}
\usepackage{amsfonts}
\usepackage{amssymb}
\usepackage{amsthm}

\newcommand{\bra}[1]{\left\langle #1 \right|}
\newcommand{\ket}[1]{\left|#1\right\rangle}
\newcommand{\braket}[2]{\left\langle#1 |  #2\right\rangle}

\newcommand\re{\mathrm{e}}
\newcommand\ri{\mathrm{i}}

\newcommand{\be}{\begin{equation}}
\newcommand{\ee}{\end{equation}}
\newcommand{\ba}{\begin{array}}
\newcommand{\ea}{\end{array}}
\newcommand{\bea}{\begin{eqnarray}}
\newcommand{\eea}{\end{eqnarray}}
\newcommand{\beas}{\begin{eqnarray*}}
\newcommand{\eeas}{\end{eqnarray*}}

\begin{document}

\title{Energy Optimal Interpolation in Quantum Evolution}
\author{Xiao Ge, Zhan Xu\\
\small Department of Physics , Tsinghua University, Beijing 100084, P.R.China\\[-0.25in]}
\date{}
\maketitle

\begin{abstract}
\noindent We introduce the concept of interpolation in quantum
evolution and present a general framework to find the energy optimal
Hamiltonian for a quantum system evolving among a given set of
middle states using variational and geometric methods. The quantum
brachistochrone problem is proved as a special case.
\end{abstract}

\section{Introduction}\label{Intro}
Recently the Quantum Brachistochrone Problem (QBP) proposed by
Carlini, et al.~\cite{Ca} has become a hot topic.\footnote{See, for
example, [2,3,4,5,6].} The aim of QBP is to find the time optimal
Hamiltonian under a given set of constraints for the quantum
evolution between two given states.

In this paper we will consider a more general problem: energy
optimal interpolation in quantum evolution, and prove that QBP is a
special case.

The interpolation in quantum evolution can be described as follows:
find a Hamiltonian in a given Hilbert space under proper
constraints, so that the quantum state $\ket{\psi(t)}$ equals given
states $\ket{\psi_1}$ ,\ldots , $\ket{\psi_m}$ at given times $t_1$
,\ldots , $t_m$ . The energy optimal interpolation (EOI) is to find
the Hamiltonian in the solution space of the former problem such
that
\begin{equation}
tr(|\hat{H}|^2)=\min. \label{1.3}\
\end{equation}

For the case m=2, the solution to EOI is the same as the result
in~\cite{Ca}.\footnote{To be proved in section \ref{QBP}.} If H is
dependent on time, the evolution between $t_i$ and $t_{i+1}$ is
reduced to the case m=2, then EOI turns trivial. In practical
consideration, the quantum evolution is usually too quick to control
the Hamiltonian for corresponding changes. Thus we only consider
time-independent Hamiltonians.

\section{General Discussion}\label{Discu}

Write the interpolation conditions in Schr\"{o}dinger equation:
\begin{equation}
    \exp({-\ri\hat{H}t_i})\ket{\psi_1} = \ket{\psi_i} , i=1,\ldots ,m, \label{2.1}\
\end{equation}
where we take $\hbar=1$ , $t_1=0$ for convenience.

A global phase factor will not alter the physical results, so
(\ref{2.1}) can be rewritten as:
\begin{equation}
    \exp({-\ri\hat{H}t_i})\ket{\psi_1} = \exp({\ri\theta_i})\ket{\psi_i} , i=1,\ldots ,m, \label{2.1*}\
\end{equation}
where $\theta_i$ can be arbitrarily adjusted to fit the physical
system.

Consider the Hilbert space of dimension n, in the eigenstate
representation the Hamiltonian is diagonalized:\footnote{In this
paper we only concern Hermitian Hamiltonians.}
\begin{equation}
    \hat{E}=\hat{T}\hat{H}\hat{T}^\dagger=diag(\varepsilon_1
    ,\ldots ,\varepsilon_n), \label{2.2}\
\end{equation}
where $\hat{T}$ is the transformation matrix,
$\hat{T}^\dagger\hat{T}=\hat{T}\hat{T}^\dagger=\hat{I}$ ,
$\varepsilon_1$ ,\ldots , $\varepsilon_n$ are the eigenvalues of
energy.

The state $\ket{\psi_i}$ now transforms to
$\ket{{\psi_i}'}=\re^{\ri\theta_i}\:\hat{T}\ket{\psi_i}$ . Write
$\hat{A}_{n\times{m}}=(\ket{{\psi_1}'},\ldots , \ket{{\psi_m}'})$ ,
then (\ref{2.1*}) becomes:
\begin{equation}
    (\lambda_k\exp({-\ri\varepsilon_k
    t_i}))_{n\times{m}}=\hat{A}, \label{2.3}\
\end{equation}
where $(\lambda_1 ,\ldots , \lambda_n)^T=\hat{T}\ket{\psi_1}$ , and
normalization requires
\begin{equation}
    \sum_{k=1}^n|\lambda_k|^2=1 \label{2.31}\
\end{equation}


Take the Hermitian adjoint of (\ref{2.3}) and then multiply it by
(\ref{2.3}) we obtain:
\begin{equation}
    (\sum_{k=1}^n|\lambda_k|^2\re^{\ri\varepsilon_k(t_i-t_j)})_{m\times{m}}
    = \hat{A}^\dagger\hat{A} =(\re^{\ri(\theta_j-\theta_i)}\braket{\psi_i}{\psi_j})_{m\times{m}},
    \label{2.4}\
\end{equation}
or writing in component form
\begin{equation}
    \sum_{k=1}^n|\lambda_k|^2\exp({\ri\varepsilon_k(t_i-t_j)+\ri(\theta_i-\theta_j)})=\Delta_{ij}
    ,1\leq{j}<i\leq{m},
    \label{2.5}\
\end{equation}
where
$\Delta_{ij}=\braket{\psi_i}{\psi_j}=\braket{\psi_j}{\psi_i}^*=\Delta_{ji}^*$
,so the case $i\leq{j}$ is trivial. This is the fundamental equation
for interpolation.

From (\ref{2.5}) we can immediately make the following observations.

\begin{itemize}
\item
    The solution to the interpolation problem
    is invariant under a unitary transformation applied to the given
    states. Hence two different sets of given states will yield the same
    result, if they are connected by a unitary transformation.


\item
    If $(\varepsilon_1,\ldots ,\varepsilon_n,\lambda_1,\ldots
    ,\lambda_n,\theta_1,\ldots ,\theta_n)$ is a solution set to the
    equations, then for arbitrary $\Delta\varepsilon$, $(\varepsilon_1+\Delta\varepsilon,\ldots ,\varepsilon_n+\Delta\varepsilon,\lambda_1,\ldots
    ,\lambda_n,\theta_1-\Delta\varepsilon t_1,\ldots
    ,\theta_n-\Delta\varepsilon t_n)$ is
    also a solution set. The zero point of energy only cause a
    total phase change, thus we can always adjust the phases to
    adjust the zero point of energy without altering the solution.

\item
    There are $m(m-1)$ equations in (\ref{2.5}) (consider the real and
    imaginary part separately). The unknowns are $\theta_i, |\lambda_k|$ and
    $\varepsilon_k$. So there are in general $m^2-m$ equations and $2n+m$ unknowns
    in total.
\end{itemize}

The structure of the solution space is determined by the number of
independent equations. When the number of equations exceeds that of
unknowns, we can use methods like least squares to find a path
optimal Hamiltonian. When the number of unknowns exceeds that of
equations, we can introduce the energy optimal Hamiltonian using
constraint (\ref{1.3}).

As
$tr(|\hat{E}|^2)=tr(\hat{T}\hat{H}^\dagger\hat{T}^\dagger\hat{T}\hat{H}\hat{T}^\dagger)=tr(|\hat{H}|)^2$
, (\ref{1.3}) now becomes
\begin{equation}
\sum_{k=1}^n\varepsilon_k^2=\min.\label{2.51}\
\end{equation}

View the $m(m-1)$ equations in (\ref{2.5}) as constraints,
(\ref{2.31}) gives another constraint. Introduce corresponding
Lagrangian multipliers $\alpha_{ij},\beta_{ij},\gamma,$ the
variational function is defined as follows:

\begin{align}
    S=\sum_{k=1}^n\varepsilon_k^2+\!\!\!\!\!\sum_{1\leq{j}<i\leq{m}}\!\!\!
    [\alpha_{ij}(\sum_{k=1}^n|\lambda_k|^2\cos({\varepsilon_k(t_i-t_j)}+\theta_i-\theta_j)-\Re\Delta_{ij})+\nonumber\\
    \beta_{ij}(\sum_{k=1}^n|\lambda_k|^2\sin({\varepsilon_k(t_i-t_j)}+\theta_i-\theta_j)-\Im\Delta_{ij})]
    +\gamma(\sum_{k=1}^n|\lambda_k|^2-1)
    \label{2.6}\
\end{align}
Variation with respect to $\varepsilon_k$ leads to:

\begin{align}
    \frac{\partial{S}}{\partial{\varepsilon_k}}= & 2\varepsilon_k-\!\!\!\!\!\sum_{1\leq{j}<i\leq{m}}\!\!\!\!
    |\lambda_k|^2(t_i-t_j)[\alpha_{ij}\sin({\varepsilon_k(t_i-t_j)+\theta_i-\theta_j})-\nonumber\\
     & \beta_{ij}\cos({\varepsilon_k(t_i-t_j)+\theta_i-\theta_j})]=0, \; k=1,\ldots ,n.
    \label{2.71}\
\end{align}
Variation with respect to $|\lambda_k|$ leads to:
\begin{align}
    \frac{\partial{S}}{\partial{|\lambda_k|}}= & 2\gamma|\lambda_k|+2|\lambda_k|\!\!\!\sum_{1\leq{j}<i\leq{m}}\!\!
    [\alpha_{ij}\cos({\varepsilon_k(t_i-t_j)+\theta_i-\theta_j})+\nonumber\\
     & \beta_{ij}\sin({\varepsilon_k(t_i-t_j)+\theta_i-\theta_j})]=0, \; k=1,\ldots ,n.
    \label{2.72}\
\end{align}
Variation with respect to $\theta_i$ leads to:
\begin{align}
    \frac{\partial{S}}{\partial{\theta_i}}= & \sum_{j\neq i} \sum_{k=1}^n|\lambda_k|^2
    [-\tilde{\alpha}_{ij}\sin({\varepsilon_k(t_i-t_j)+\theta_i-\theta_j})+\nonumber\\
     & \tilde{\beta}_{ij}\cos({\varepsilon_k(t_i-t_j)+\theta_i-\theta_j})]=0, \;i=1,\ldots ,m.
    \label{2.73}\
\end{align}
where
\begin{align}
\tilde{\alpha}_{ij}= \left\{ \begin{array}{ll}
\alpha_{ij} & i>j,\\
\alpha_{ji} & i<j,\end{array} \right. \;\; \tilde{\beta}_{ij}=
\left\{\begin{array}{ll}
\beta_{ij} & i>j,\\
\beta_{ji} & i<j.\end{array} \right.
\end{align}

Hence the fundamental equations for EOI are (\ref{2.5}),
(\ref{2.71}), (\ref{2.72}) and (\ref{2.73}). The unknowns are
$\varepsilon_k$, $|\lambda_k|$, $\theta_i$, $\alpha_{ij}$,
$\beta_{ij}$ and $\gamma$. In total there are $m^2+2n+1$ equations
and $m^2+2n+1$ unknowns, and can be solved in principle.



\section{Quantum Brachistochrone Problem}\label{QBP}

Constraint (\ref{1.3}) is to minimize the module of $H$ while the
evolution time keep fixed. From Anandan-Aharonov relation~\cite{AA},
which states that the `speed' of quantum evolution is given by
$2\Delta \hat{H}/\hbar$ , $\Delta \hat{H}$ being the standard
deviation of the Hamiltonian, we conclude the reversed constraint ,
viz.
\begin{equation}
    tr(|\hat{H}|^2)=\textrm{const}.\; , t=\min.\;, \label{1.10}\
\end{equation}
should be equivalent to (\ref{1.3}).

Carlini, et al.~\cite{Ca} used the constraint
\begin{equation}
    tr(|\tilde{H}|^2)=tr(|\hat{H}-tr(\hat{H})/n|^2)=\textrm{const}.\;
    , t=\min.\; , \label{1.11}\
\end{equation}
in solving QBP. The difference of $\tilde{H}$ and $\hat{H}$ only
lies in the choice of the zero point of energy.

From (\ref{2.5}) we know the the choice of the zero point of energy
would not alter the solution, since
$tr(|\tilde{H}|^2)=tr(|\hat{H}|^2)-|tr(\hat{H})|^2/n\leq{tr(|\hat{H}|^2)}$,
constraint (\ref{1.3}) requires
\begin{equation}
    tr(\hat{H})=0, \label{1.12}\
\end{equation}
i.e. $tr(|\tilde{H}|^2)=tr(|\hat{H}|^2)$. Hence (\ref{1.10}) and
(\ref{1.11}) are equivalent. So the the special case $m=2$ for EOI
should reduce to QBP. Now (\ref{2.5}) and (\ref{2.71}) become:

\be
\begin{aligned}
    2\varepsilon_k= & |\lambda_k|^2\,t\,[\alpha\sin({\varepsilon_kt+\theta})-
    \beta\cos({\varepsilon_kt+\theta})]\;,\; k=1,\ldots ,n;\\
     & \sum_{k=1}^n|\lambda_k|^2\exp(\ri({\varepsilon_kt+\theta}))=\Delta\;,
    \label{3.1}\
\end{aligned}
\ee
for convenience we set $t_1=\theta_1=0$ and omit redundant
subscripts. Treat $\theta_i$'s and $|\lambda_k|$'s as known
parameters we can solve (\ref{3.1}), then adjust the values of
$\theta_i$'s and $|\lambda_k|$'s to minimize the energy cost.

To simplify (\ref{3.1}), we shall now consider from the geometry
viewpoint. The two given states $\ket{\psi_i}$ and $\ket{\psi_f}$
span a space of dimension 2. It's natural that the time optimal
evolution should be the geodesic in this space. So the Hamiltonian
is of dimension 2.\footnote{See~\cite{BH} for a proof.}\footnote{We
suppose the optimal Hamiltonian for m given independent states is of
dimension m, which is a unproved guess.}

From (\ref{1.12}) we have
\begin{equation}
    \sum_{k=1}^n\varepsilon_k=tr(\hat{E})=tr(\hat{H})=0.
    \label{3.1a}\
\end{equation}
Thus
\begin{equation}
    \varepsilon_1=-\varepsilon_2=\varepsilon. \;\;\;\varepsilon_k=0,
    k=3,\ldots ,n.
    \label{3.2}\
\end{equation}
Then (\ref{3.1}) reduce to:
\be
\begin{aligned}
    2\varepsilon &=|\lambda_1|^2\,t\,[\alpha\sin({\varepsilon t+\theta})-
    \beta\cos({\varepsilon t+\theta})]; \\
    -2\varepsilon &=|\lambda_2|^2\,t\,[\alpha\sin({-\varepsilon t+\theta})-
    \beta\cos({-\varepsilon t+\theta})]; \\
    0 &=|\lambda_k|^2\,t\,[\alpha\sin{\theta}-\beta\cos{\theta}], k=3,\ldots ,n;\\
    \Delta\re^{-\ri\theta} &=|\lambda_1|^2\re^{\ri\varepsilon t}+
    |\lambda_2|^2\re^{-\ri\varepsilon t}+
    \sum_{k=3}^n|\lambda_k|^2. \label{3.3}\
\end{aligned}
\ee

The third eqn. of (\ref{3.3}) yields $|\lambda_k|=0, k\geq3.$
Substitute it into the forth eqn. we have:
\begin{equation}
    \Lambda\re^{\ri\varepsilon t}+(1-\Lambda)\re^{-\ri\varepsilon
    t}=\Delta\re^{-\ri\theta},
    \label{3.4}\
\end{equation}
where we have used (\ref{2.31}) and write $|\lambda_1|^2=\Lambda$
for convenience.

Take module on both sides of (\ref{3.4}) :
\begin{equation}\varepsilon t
    |\Delta|^2
    =1-4\Lambda(1-\Lambda)\sin(\varepsilon
    t)^2.
    \label{3.5}\
\end{equation}

Thus
$\varepsilon=\arcsin{\sqrt{\frac{1-|\Delta|^2}{4\Lambda(1-\Lambda)}}}/t$,
to minimize $\varepsilon$, we take $\Lambda=1/2$. So
\begin{equation}
\varepsilon=\arcsin{\sqrt{1-|\Delta|^2}}/t=\arccos{|\Delta|}/t,
\label{3.51}\
\end{equation}
\begin{equation}
    tr(|\hat{H}|^2)=\sum_{k=1}^n{\varepsilon_k^2}=2(\arccos{|\Delta|}/t)^2.
    \label{3.6}\
\end{equation}

In the eigenstate representation,
$\hat{H}=diag(\varepsilon,-\varepsilon,0,\ldots,0),
\ket{\psi_1}=(\lambda_1,\lambda_2,0,\ldots,0)^T,\\
\ket{\psi_2}=\exp(-\ri\hat{H}t)\ket{\psi1}=(\lambda_1\exp(-\ri\varepsilon
t),\lambda_2\exp(\ri\varepsilon t),0,\ldots,0)^T.$ The Gram-Schmidt
orthonormalized state $\ket{\psi'_2}=(\ket{\psi_2}-\cos\varepsilon
t\ket{\psi_1})/\sin\varepsilon
t=(-\ri\lambda_1,\ri\lambda_2,0,\ldots,0)^T$, so \be
    \ri\varepsilon(\ket{\psi'_2}\bra{\psi_1}-\ket{\psi_1}\bra{\psi_2'})
    =diag(\varepsilon,-\varepsilon,0,\ldots,0)=\hat{H}.\label{3.61}\
\ee
At arbitrary time $\tau$, \be
    \ket{\psi(\tau)}=(\lambda_1\exp(-\ri\varepsilon\tau),
    \lambda_2\exp(\ri\varepsilon\tau),0,\ldots,0)^T=
    \cos\varepsilon\tau\ket{\psi_1}+\sin\varepsilon\tau\ket{\psi_2'}.
    \label{3.62}\
\ee
(\ref{3.6}),(\ref{3.61}) and (\ref{3.62}) are the same as the
main results in~\cite{Ca} which take $tr(|\hat{H}|^2)/2=\omega^2$.

Substitute the results back into (\ref{3.3}) we have:
\begin{equation}
    |\Delta|=\Delta\text{e}^{-\ri\theta} , \label{3.7}\
\end{equation}
thus $\braket{\psi_2}{\psi_1}=\Delta\text{e}^{-\ri\theta}$ is real.
The global phases of $\psi_1$ and $\psi_2$ are adjusted to make
$\braket{\psi_2}{\psi_1}$ be a real number at optimal solution.

\section{Case $\Delta_{ij}=\delta_{ij},t_i=(i-1)t$}
Now we consider the special case when the given states are
orthogonal and the given times are evenly spaced, viz.
\begin{equation}
    \Delta_{ij}=\delta_{ij}\;,\;t_i=(i-1)t.\label{4.1}\
\end{equation}
Since the given $\ket{\psi_i}$'s are lineal independent. We have:
\begin{equation}
    m\leq n \label{4.2}\
\end{equation}

Now (\ref{2.5}) reduces to:
\begin{equation}
    \sum_{k=1}^n|\lambda_k|^2\exp({\ri(l\varepsilon_k t+\theta_l)})=0
    ,\;l=i-j=1,2,\ldots,m-1.
    \label{4.3}\
\end{equation}

In section \ref{QBP} we concluded that the global phases $\theta_i$
are adjusted to make $\Delta_{ij}$ real at optimal solution. This
conclusion is supposed to be still valid here. Thus we have
$\theta_l=0$.

When $l=1$ a special solution can be easily found: the n vectors
$\exp({i\varepsilon_k t}) (k=1,\ldots,n)$ distribute uniformly on
the unit circle, i.e.
\begin{equation}
    \varepsilon_k
    t=\frac{2k\pi}{n}+\theta_0,\;|\lambda_k|^2=1/n,
    \label{4.4}\
\end{equation}

If n is prime, for any $l<n$, $\exp({il\varepsilon_k t})$ are still
n vectors distributing uniformly on the unit circle. If n is not
prime, when l is a factor of n, $\exp({il\varepsilon_k t})$ are
$n/l$ vectors distributing uniformly on the unit circle. In both
cases we still have (\ref{4.3}). But if $l=n$, (\ref{4.3}) no longer
holds, this is consistent with (\ref{4.2}).

The result is a special solution to the quantum interpolation,
whether it is the solution to EOI depends on whether it is
consistent with (\ref{2.71}), (\ref{2.72}) and (\ref{2.73}), which
remains unproved.

%

\section{Conclusion and Discussion}

We have educed the fundamental equations for the quantum
interpolation and EOI. The general behavior of its solutions are
preliminarily studied. QBP is discussed in detail as a special case.
Another simple case for quantum interpolation is also considered.

As future developments, the fundamental equations for EOI remains to
be thoroughly investigated, and this framework can be extended to
mixed states and non-Hermitian Hamiltonians. The behavior of path
optimal Hamiltonian\footnote{See page \pageref{2.51}, probably
defined via methods such as least squares.} is also worth studying.
Meanwhile applications in quantum computation are probable.

\section*{Acknowledgments}
The idea of this paper was inspired by the seminar on quantum
mechanics by Prof. Zhuang, Pengfei. And we would like to thank
Huang, Xuguang, our teaching assistant, for informative discussions.

\end{document}